\documentclass[aps, prd, reprint, superscriptaddress]{revtex4-2}
\usepackage{graphicx}
\usepackage{bm}
\usepackage{amsmath}
\usepackage{amsfonts}
\usepackage{amssymb}
\newcommand{\bra}[1]{\left\langle #1 \right\vert}
\newcommand{\ket}[1]{\left\vert #1 \right\rangle}

\newcommand{\abs}[1]{\left\vert #1 \right\vert}

\begin{document}

\title{Use VQE to calculate the ground energy of hydrogen molecules on IBM Quantum}

\author{Maomin Qing}
\email{kands-code@qq.com}
\affiliation{College of Science, China Three Gorges University, Yichang 443002, People's Republic of China}
\affiliation{Center for Astronomy and Space Sciences, China Three Gorges University, Yichang 443002, People's Republic of China}

\author{Wei Xie}
\email{xiewei@ctgu.edu.cn}
\affiliation{College of Science, China Three Gorges University, Yichang 443002, People's Republic of China}
\affiliation{Center for Astronomy and Space Sciences, China Three Gorges University, Yichang 443002, People's Republic of China}

\date{\today}

\begin{abstract}
  Quantum computing has emerged as a promising technology for solving problems that are intractable for classical computers.
  In this study, we introduce quantum computing and implement the Variational Quantum Eigensolver (VQE) algorithm using Qiskit on the IBM Quantum platform to calculate the ground state energy of a hydrogen molecule.
  We provide a theoretical framework of quantum mechanics, qubits, quantum gates, and the VQE algorithm.
  Our implementation process is described, and we simulate the results.
  Additionally, experiments are conducted on the IBM Quantum platform, and the results are analyzed.
  Our findings demonstrate that VQE can efficiently calculate molecular properties with high accuracy.
  However, limitations and challenges in scaling the algorithm for larger molecules are also identified.
  This work contributes to the growing body of research on quantum computing and highlights the potential applications of VQE for real-world problem-solving.
\end{abstract}

\maketitle

\section{\label{sec:introduction}Introduction}

Quantum computing is a rapidly growing field that explores the potential of quantum mechanics to develop new technologies\cite{quantcomp}.
Unlike classical computers which use binary digits (bits) to represent information, quantum computers use quantum bits (qubits) that can exist in superpositions of states, allowing for massive parallelism and the ability to solve problems faster than classical computers.

In this study, we focus on the implementation of the Variational Quantum Eigensolver (VQE) algorithm\cite{vqe} using Qiskit\cite{qiskit} on the IBM Quantum platform\cite{ibmquan} to calculate the ground state energy of a hydrogen molecule.
The calculation of molecular properties is of great significance in chemistry, as it can aid in the design of new drugs, catalysts, and materials.
However, the computational complexity of solving the Schr{\"o}dinger equation for large molecules using classical methods limits the accuracy and efficiency of these calculations.
VQE has emerged as a promising algorithm for calculating molecular energies on quantum computers, offering significant speed-ups over classical methods.

Quantum computing is an emerging field of computer science that utilizes the principles of quantum mechanics to perform calculations on data.
In traditional computing, bits are used to represent information in binary form (0 or 1), whereas in quantum computing, qubits are used to store and manipulate information.
Qubits are quantum systems that can exist in a superposition of states, which means they can represent both 0 and 1 at the same time.
This property makes quantum computers capable of performing certain calculations exponentially faster than classical computers.

One of the most famous algorithms in quantum computing is Shor's algorithm\cite{shor}, which efficiently factors large numbers into their prime factors.
This algorithm has important implications for cryptography, as it would allow for the efficient breaking of many commonly used encryption schemes.
Another important algorithm in quantum computing is Grover's algorithm\cite{grover}, which provides a quadratic speedup for searching unstructured databases.
Grover's algorithm has numerous applications in fields such as optimization and machine learning.

The basic building block of a quantum computer is a quantum gate, which is a unitary transformation that acts on one or more qubits.
Some common examples of quantum gates include the Hadamard gate, which creates a superposition state, and the Pauli gates, which perform rotations around the x, y, and z axes.
The state of a quantum system is represented by a ket vector, which is an element of a complex Hilbert space.
In order to perform quantum computations, it is necessary to perform operations on qubits while maintaining their coherence.
This is a significant challenge due to the susceptibility of quantum systems to decoherence from environmental interactions.
Despite these challenges, there has been significant progress in the development of quantum hardware and algorithms in recent years.
Many companies and research institutions are actively working on developing practical quantum computers that can be used to solve real-world problems.

The VQE algorithm is a hybrid quantum-classical approach for finding the ground state energy of a given molecule on a quantum computer.
The goal of the VQE algorithm is to find the lowest eigenvalue of the molecular Hamiltonian, which represents the ground energy of the molecule.
In essence, the VQE algorithm involves preparing a trial wavefunction called Ansatz on the quantum computer, measuring its energy, and then optimizing the parameters of the wavefunction using a classical optimization algorithm.
This process is repeated iteratively until an optimal set of parameters is found that minimizes the energy of the trial wavefunction.

Given a molecular Hamiltonian $\hat{H}$, we can express the ground state energy $E_0$ as the minimum of the expectation value of $\hat{H}$ with respect to some trial wavefunction $\ket{\psi(\theta)}$ parametrized by a set of parameters $\theta$:

\begin{equation}
  E_0 = \min_\theta \bra{\psi(\theta)} \hat{H} \ket{\psi(\theta)}
\end{equation}

We can use a quantum computer to measure $\bra{\psi} \hat{H} \ket{\psi}$, and a classical optimizer to vary the parameters $\theta$ in order to minimize this expectation value.
This process is repeated until convergence is achieved and an approximation of the ground state energy is obtained.

This work will begin with an introduction to quantum computing and explain the principle behind the VQE algorithm.
In addition, we utilized IBM's Quantum platform and Qiskit library to calculate the ground-state energy of a hydrogen molecule using the VQE algorithm for our research.
Finally, we will summarize the current development direction of quantum algorithms and provide future outlooks.

\section{\label{sec:quantumcomputing}Quantum Computing}

\subsection{\label{subsec:representofqubit}Represent of Qubits}

In quantum computing, wave functions and qubits are usually represented using Dirac notation, which is written in the form of a Ket vector.
Typically, the state of a qubit can be represented by a ket vector as follow:

\begin{equation}
  \begin{gathered}
    \ket{\psi} = \alpha \ket{0} + \beta \ket{1} = \begin{bmatrix}
      \alpha \\ \beta
    \end{bmatrix} \\
    \ket{0} = \begin{bmatrix}
      1 \\ 0
    \end{bmatrix},
    \ket{1} = \begin{bmatrix}
      0 \\ 1
    \end{bmatrix}
  \end{gathered}
\end{equation}

The $\alpha$ and $\beta$ are called probability amplitude, satisfy the normalization of probability, means $\abs{\alpha}^2 + \abs{\beta}^2 = 1$.
The $\ket{0}$ and $\ket{1}$ are called standard basis vectors, correspond to 0 and 1 respectively.

\begin{figure}[!ht]
  \centering
  \includegraphics[scale=0.4]{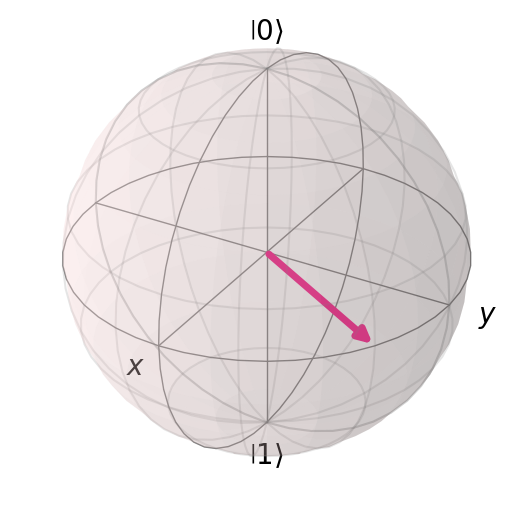}
  \caption{An example of a state vector represented in a Bloch sphere}
  \label{fig:bloch-example}
\end{figure}

Qubits can also be represented in three-dimensional space by Bloch spheres, as shown in FIG. \ref{fig:bloch-example}.

At this time, since the length of the vector is specified as a unit length, the quantum state can be determined by the angle $\phi$ between the projection of the vector on the $x$-$y$ plane and the $x$ axis and the angle $\theta$ between the vector and the $z$ axis.
At this time, the state vector can be expressed as:

\begin{equation}
  \ket{\psi} = \cos{\dfrac{\theta}{2}} \ket{0}
  + e^{i \phi} \sin{\dfrac{\theta}{2}} \ket{1}
\end{equation}

One of the benefits of this representation is the ability to distinguish global phase.
The so-called global phase is a factor whose norm is 1.

The state of a system composed of $n$ qubits is represented by the tensor product composite of the states of these $n$ qubits, that is, it can be represented by a $2^n$-dimensional complex vector, where each element corresponds to a ground state composed of a qubit.
For example, in a system of two qubits, if the first qubit is in state $\ket{0}$ and the second qubit is in state $\ket{1}$, the state of the entire system can be expressed as:

\begin{equation}
  \ket{\psi} = \ket{1} \otimes \ket{0} = \ket{1 0} = \begin{bmatrix}
    0 \\ 0 \\ 1 \\ 0
  \end{bmatrix} \\
\end{equation}

Similar to the single-qubit case, we can also use the Dirac notation to represent the state of a multi-qubit system, namely:
\begin{equation}
  \ket{\psi} = \sum_{i = 0}^{2^n - 1}{c_i \ket{i}_n}
\end{equation}
where $c_i$ is a complex coefficient, and $\ket{i}_n$ denotes the $i + 1$ th ground state in a system of $n$ qubits.
In quantum computing, we usually only focus on the states associated with a certain measurement, i.e. if we measure the system, it will only be in one of the ground states.

\subsection{Operations on Qubits}

In quantum computing, all operations can be equivalently performed by applying a unitary matrix to a qubit.

For example, if you want to flip the state vectors of qubits around their respective axes, that is, rotate $\pi$ around the corresponding axis, this operation corresponds to the Pauli matrix.
In quantum computing, Pauli matrices are also known as Pauli gates.

\begin{equation}
  X = \begin{bmatrix}
    0 & 1 \\ 1 & 0
  \end{bmatrix},
  Y = \begin{bmatrix}
    0 & -i \\ i & 0
  \end{bmatrix},
  Z = \begin{bmatrix}
    1 & 0 \\ 0 & -1
  \end{bmatrix}
\end{equation}

Generally, the identity matrix $I$ is regarded as a kind of Pauli gate.
Compared with the Pauli gate, there is also a special gate that rotates around a certain axis, called the rotation gates, denoted as $R_k(\theta)$, where $k$ represents the corresponding axis, and $\theta$ represents the rotation angle.

\begin{equation}
  \label{eq:routation-gate}
  \begin{aligned}
    R_x(\theta)
    = \exp(- i X \theta / 2)
     & = \begin{bmatrix}
           \cos(\theta / 2)     & - i \sin(\theta / 2) \\
           - i \sin(\theta / 2) & \cos(\theta / 2)
         \end{bmatrix} \\
    R_y(\theta)
    = \exp(- i Y \theta / 2)
     & = \begin{bmatrix}
           \cos(\theta / 2) & - \sin(\theta / 2) \\
           \sin(\theta / 2) & \cos(\theta / 2)
         \end{bmatrix}       \\
    R_z(\theta)
    = \exp(- i Z \theta / 2)
     & = \begin{bmatrix}
           \exp(- i \theta / 2) & 0                  \\
           0                    & \exp(i \theta / 2)
         \end{bmatrix}
  \end{aligned}
\end{equation}

According to Euler's formula $\exp(i \pi) + 1 = 0$, the $Z$ gate can be squared to obtain the $S$ gate, and the $S$ gate can also be squared to obtain the $T$ gate.
Correspondingly, the $S$ gate rotates $\pi/2$ around the $z$-axis, while the $T$ gate rotates $\pi/4$.

\begin{equation}
  \begin{aligned}
    S & = Z^{1/2}                                       \\
      & = \begin{bmatrix}
            1 & 0 \\ 0 & i
          \end{bmatrix} = \exp(i \pi / 4) R_z(\pi / 2)  \\
    T & = S^{1/2} = Z^{1/4}                             \\
      & = \begin{bmatrix}
            1 & 0 \\ 0 & \exp(i \pi / 4)
          \end{bmatrix} = \exp(i \pi / 8 ) R_z(\pi / 4)
  \end{aligned}
\end{equation}

In quantum computing, there is another common operation, which is to prepare a uniformly distributed superposition state.
The so-called uniform distribution superposition state is a quantum superposition state, where $\ket{0}$ and $\ket{1}$ have the same probability, that is, $(1/\sqrt{2})(\ket{0} + \ket{1})$ and $(1/\sqrt{2})(\ket{0} - \ket{1})$.
These two states can usually be represented by $\ket{+}$ and $\ket{-}$.

A uniform superposition state is usually achieved using a Hadamard gate, denoted as $H$.
It introduces transitions between uniformly distributed superposition states and non-uniformly distributed states.
The so-called non-uniformly distributed states are $\ket{0}$ and $\ket{1}$ for a qubit.

\begin{equation}
  \begin{gathered}
    H = \dfrac{1}{\sqrt{2}}\begin{bmatrix}
      1 & 1 \\ 1 & -1
    \end{bmatrix} \\
    H \ket{0} = \ket{+},\ H \ket{1} = \ket{-} \\
    H \ket{+} = \ket{0},\ H \ket{-} = \ket{1}
  \end{gathered}
\end{equation}

For multi-qubit systems, the more common quantum gate is the controlled NOT gate (CNOT).
CNOT is divided into two parts, the control qubit and the target qubit.
If the state of the control qubit is $\ket{1}$, it is equivalent to applying an $X$ gate to the target qubit, otherwise it will do nothing to the target qubit.

\begin{equation}
  \begin{gathered}
    \textnormal{CNOT} = \begin{bmatrix}
      1 & 0 & 0 & 0 \\ 0 & 1 & 0 & 0 \\ 0 & 0 & 0 & 1 \\ 0 & 0 & 1 & 0
    \end{bmatrix} \\
    \textnormal{CNOT} \ket{10} = \ket{11},\ \textnormal{CNOT} \ket{11} = \ket{10} \\
    \textnormal{CNOT} \ket{--} = \ket{+-},\ \textnormal{CNOT} \ket{+-} = \ket{--}
  \end{gathered}
\end{equation}

A set of quantum gates is said to be universal if there is a set of quantum gates such that all quantum operations, or quantum programs, can be approximated by sequences of gates in this set.
Any quantum program can be represented by a sequence of quantum circuits and classical near-time computation.

A more common universal gate set is $\{ \textnormal{CNOT}, SGs \}$\cite{universalgate01},  where $SGs$ represents all signle-qubit gates.
Next is $\{ \textnormal{CNOT}, H, T \}$\cite{universalgate02}.

\subsection{Quantum Circuits}

In quantum computing, quantum circuits are graphical representations used to describe the evolution and manipulation of quantum states.
It consists of a series of quantum gates, each of which represents an operation on one or more qubits.
By constructing different quantum circuits, we can realize tasks such as various quantum algorithms and quantum communication protocols\cite{universalgate01}.

\begin{figure}[ht!]
  \centering
  \includegraphics[scale=0.96]{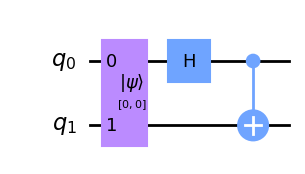}
  \caption{An example shows that use $H$ gate and CNOT gate to prepare entagled states}
  \label{fig:multi-entaglement}
\end{figure}

For example, the FIG. \ref{fig:multi-entaglement} shows the process of preparing entangled states using $H$ gate and CNOT gate.

\section{\label{sec:vqehydro}Use VQE to get hydrogen molecule}

The VQE algorithm is a quantum algorithm for solving the smallest eigenvalue, and is usually used to solve the molecular ground state energy.
The core idea of the VQE algorithm is to construct a parameterized quantum circuit, and then use a classical computer to optimize the quantum circuit so that the output state of the circuit is close to the target state, and calculate the expected value of the target Hamiltonian on this state.
By minimizing this expected value, the minimum eigenvalue of the target Hamiltonian can be obtained, and then the solution to the molecular ground state energy can be completed.

Before starting to solve the problem, we first organize the Hamiltonian and wave function of the hydrogen atom into the form required in quantum computing, that is, the form represented by the combination of quantum gates.

In quantum mechanics, for a quantum state $\ket{\psi}$, we can use the Hamiltonian operator to replace its energy.
In other words, the system energy is the eigenvalue of its Hamiltonian, and the ground state energy is its minimum eigenvalue.

\begin{equation}
  \begin{gathered}
    \hat{H} \ket{\psi} = E \ket{\psi} \implies \bra{\psi} \hat{H} \ket{\psi} = E \\
    \bra{\psi} \hat{H} \ket{\psi} = E \ge E_{\textnormal{min}}
  \end{gathered}
\end{equation}

In this way, our goal becomes to find the minimum value of $\bra{\psi} \hat{H} \ket{\psi}$, that is, to find $\min_\theta \bra{\psi(\theta)} \hat{H} \ket{\psi(\theta)}$.

\subsection{The Hamiltonian}

The Hamiltonian of a system can generally be split into two parts: momentum $\hat{T}$ and potential energy $\hat{V}$.
where $\hat{V}$ can usually be expressed as $V(r)$, which is a function of distance.
In an electronic system, the potential energy can be roughly divided into three parts, electrons and electrons, nucleus and nuclei, and electrons and nuclei, which can be expressed as follows:
\begin{equation}
  \begin{aligned}
    \hat{V} = V(r) & = \sum_{i, j}^{\textnormal{electrons}}{\dfrac{e^2}{4 \pi \epsilon_0 \abs{r_i - r_j}}} + \sum_{i, j}^{\textnormal{nuclei}}{\dfrac{Z_i Z_j e^2}{4 \pi \epsilon_0 \abs{r_1 - r_j}}} \\
                   & - \sum_{i}^{\textnormal{electrons}}\sum_{j}^{\textnormal{nuclei}}\dfrac{Z_j e^2}{4 \pi \epsilon_0 \abs{r_i - r_j}}
  \end{aligned}
\end{equation}
where $Z_j$ is the number of protons in the $j$-th nucleus, $e$ is the electronic charge.

Kinetic energy can be divided into two parts: electron kinetic energy and atomic nuclear kinetic energy.
\begin{equation}
  \hat{T} =  - \sum_i^{\textnormal{nuclei}} \dfrac{\hbar^2}{2 m_i} \nabla^2_i - \sum_i^{\textnormal{electrons}}\dfrac{\hbar^2}{2 m_e} \nabla^2_i
\end{equation}
where $m_i$ is the mass of a nucleus, $m_e$ is the electron mass, $\nabla^2_i$ is the Laplace operator of the $i$-th particle.

For these two complex formulas, we can simplify them.
First, we can assume that the nucleus is stationary, but this is relative to the electrons.
Secondly, we use atomic units\cite{atomunit} and set some constants such as $\hbar^2/m_e$ to 1, so that our formula can be simplified to the following form.

\begin{equation}
  \begin{aligned}
    \hat{H} & \approx -\sum_i^{\textnormal{electrons}}\dfrac{1}{2}\nabla^2_i + \sum_{i, j}^{\textnormal{electrons}}\dfrac{1}{\abs{r_i - r_j}} \\
            & - \sum_i^{\textnormal{electrons}} \sum_j^{\textnormal{nuclei}} \dfrac{Z_j}{\abs{r_i - r_j}} + C^{\prime}_n
  \end{aligned}
\end{equation}

Further, we can divide the Hamiltonian into two parts, the one-particle part $\sum_i h_1$ and the two-particle part $\sum_i h_2$.
Assuming that $\psi_j$ represents the spin-orbit that constitutes the system, we will use the generation operator and annihilation operator to represent the single-particle part.

\begin{equation}
  \begin{gathered}
    \sum_i h_1(x_i) = \sum_{p,q} \bra{p} \hat{H} \ket{q} \hat{a}^{\dagger}_p \hat{a}_q \\
    \begin{aligned}
       & \bra{p} \hat{H} \ket{q}                       \\
       & = \int_{-\infty}^{\infty}{\psi^{\star}_p(x_i)
      - \dfrac{1}{2}\nabla_i^2 + \sum_j{\dfrac{Z_j}{\abs{r_i - r_j}}}
      \psi_q(x_i) \mathrm{d}{x_i}}                     \\
       & = h_{p q}
    \end{aligned}
  \end{gathered}
\end{equation}

The two-particle part can also be similarly expressed by the production operator and annihilation operator.

\begin{equation}
  \begin{gathered}
    \sum_{i,j}{h_2(x_i, x_j)}
    = \sum_{p,q,r,s}{\bra{p q} \hat{H} \ket{r s}
      \hat{a}^{\dagger}_p \hat{a}^{\dagger}_q
      \hat{a}_r \hat{a}_s} \\
    \begin{aligned}
       & \bra{p q} \hat{H} \ket{r s}                        \\
       & = \int_{-\infty}^{\infty}\int_{-\infty}^{\infty}
      \psi_p^{\star}(x_i)\psi_q^{\star}(x_j)
      \dfrac{1}{\abs{x_1 - x_2}}
      \psi_r(x_j) \psi_s(x_i)\mathrm{d}{x_i}\mathrm{d}{x_j} \\
       & = h_{pqrs}
    \end{aligned}
  \end{gathered}
\end{equation}

As mentioned above, we can get the final representation of our quadratic quantized Hamiltonian.
\begin{equation}
  \label{eq:secondaryquan}
  \hat{H} = \sum_{p,q}h_{pq} \hat{a}^{\dagger}_p\hat{a}_q + \dfrac{1}{2}\sum_{p,q,r,s}{h_{pqrs}\hat{a}^{\dagger}_p\hat{a}^{\dagger}\hat{a}_r\hat{a}_s} + h_0
\end{equation}
where $h_0$ is a correction constant used to correct errors due to simplification.

Next, we also need to map the twice-quantized Hamiltonian into the quantum computer.
Here, we have many options, such as using Jordan-Wigner transformation\cite{jwtrans}, or Bravyi-Kitaev transformation\cite{bktrans}.
For simplicity, we use the Jordan-Wigner transform.
\begin{equation}
  \label{eq:jwtrans}
  \begin{gathered}
    a^{\dagger}_n \mapsto \dfrac{1}{2} \left[ \prod_{j=0}^{n-1} -Z_j \right] (X_n - i Y_n) \\
    a_n \mapsto \dfrac{1}{2} \left[ \prod_{j=0}^{n-1} -Z_j \right] (X_n + i Y_n) \\
  \end{gathered}
\end{equation}
where $X_n$, $Y_n$ and $Z_n$ are Pauli matrices in quantum computing, respectively.

After doing this, the Hamiltonian of our hydrogen molecule can be expressed as:
\begin{equation}
  \label{eq:hydrogen}
  \begin{aligned}
    \hat{H} & = -\frac{1}{2}\left(\hat{I}\otimes\hat{I}
    + \hat{X}\otimes\hat{X}
    + \hat{Y}\otimes\hat{Y}
    + \hat{Z}\otimes\hat{Z}\right)                                           \\
            & + d\left(\hat{Z}\otimes\hat{I} + \hat{I}\otimes\hat{Z}\right),
  \end{aligned}
\end{equation}

\subsection{The Wave Function}

In computational physics, usually, the wave function of our system can be expressed in the form of Slater determinant\cite{slater}.
\begin{equation}
  \psi_{(x_1, x_2, \cdots ,x_n)} = \dfrac{1}{\sqrt{N !}}
  \begin{vmatrix}
    \chi_i(x_1) & \chi_j(x_1) & \cdots & \chi_k(x_1) \\
    \chi_i(x_2) & \chi_j(x_2) & \cdots & \chi_k(x_2) \\
    \vdots      & \vdots      & \ddots & \vdots      \\
    \chi_i(x_n) & \chi_j(x_n) & \cdots & \chi_k(x_n)
  \end{vmatrix}
\end{equation}
where $1 / \sqrt{N !}$ is the normalization parameter and $\chi_m(x_l)$ represents the molecular orbital wavefunction of the $m$th orbital of the $l$th atom.
Because it basically satisfies the format of quantum computing, sometimes we can express it as $\ket{\chi_1 \cdots \chi_k}$.

But in order to make our calculation more convenient, we can quantize this wave function twice and express it in the form of occupation number representation.
\begin{equation}
  \ket{\psi} = \sum{C_i \ket{n_1 n_2 \cdots n_m}}
\end{equation}
Among them, $C_i$ is a constant coefficient, while $n_i$ represents the number of possession, and $n_i \in \{0, 1\}$ indicates whether the electron is in a certain state.

For the wave function represented by the representation of the occupation number, we use the generation operator $\hat{a}^{\dagger}_i$ and the annihilation operator $\hat{a}_i$ to operate, that is, flip the corresponding occupancy number state, the generation operator can change the number of possession from $0$ to $1$, and the annihilation operator can change the number of possession from $1$ to $0$.

\section{Use Qiskit to Solve}

Theoretically, we need to prepare the objective function and quantum circuit by ourselves, that is, use the formula \eqref{eq:secondaryquan} and the formula \eqref{eq:jwtrans} to perform quadratic Quantize and express it in the form of a quantum circuit.
For example, for hydrogen molecule, its Hamiltonian can be represented by the formula \eqref{eq:hydrogen}.

However, in the process of actually calculating the molecular ground state energy, we do not need to manually solve and simplify the molecular Hamiltonian.
In chemistry, scientists already store some common molecular properties in some databases.
When we calculate, in most cases, we only need to query the information of the corresponding molecule to get what we need.

In Python, there is a library named PySCF\cite{pyscf}, which stores the information of many elements and molecules.
We can use the drivers or APIs provided by PySCF to query molecular information or build molecules relatively easily.
For example, if we want to build a hydrogen molecule here, we can use the code showing in TABLE. \ref{code:buildhydrogen} to complete it.

\begin{table}[ht!]
  \vspace*{1em}
  \centering
  \begin{verbatim}
  molecule = Molecule(
    geometry=[
      ["H", [0.0, 0.0, -dist / 2]],
      ["H", [0.0, 0.0, dist / 2]]],
    multiplicity=1, charge=0)
  driver = ElectronicStructureMoleculeDriver(
    molecule=molecule,
    basis="sto3g",
    driver_type=ElectronicStructureDriverType.PYSCF)
  problem = ElectronicStructureProblem(driver,
    [electronic.FreezeCoreTransformer(
      freeze_core=True)])  
  \end{verbatim}
  \vspace*{-1.5em}
  \caption{Building Hydrogen Molecules Using PySCF}
  \label{code:buildhydrogen}
\end{table}

After building the model, we also need to perform secondary quantization and mapping.
These operations have been defined in Qiskit.
We can use the \verb|second_q_ops| method to obtain the operator after the second quantization.
Among the returned operators, the first one is the Hamiltonian we need, that is, \verb|hamiltonian = problem.second_q_ops()[0]|.

Secondary quantization is not enough, we also need to map it to a form that can be represented in quantum computing, here we can directly use Jordan Wigner mapping to complete this step:

\begin{table}[ht!]
  \vspace*{1em}
  \centering
  \begin{verbatim}
  mapper = JordanWignerMapper()
  converter = QubitConverter(mapper,
    two_qubit_reduction=False)
  qubit_op = converter.convert(hamiltonian)  
  \end{verbatim}
  \vspace*{-1.5em}
  \caption{Hydrogen Molecules and Hamiltonians Using Jordan Wigner Mapping}
\end{table}

After processing, the Hamiltonian can be represented using a Pauli matrix:

\begin{equation}\label{eq:realHamilton}
  \begin{aligned}
    H = & - 0.807184\ I \otimes I \otimes I \otimes I + 0.175106\ Z \otimes I \otimes Z \otimes I \\
        & + 0.169404\ I \otimes Z \otimes I \otimes Z                                             \\
        & - 0.230474\ (I \otimes I \otimes Z \otimes I + Z \otimes I \otimes I \otimes I)         \\
        & + 0.173740\ (I \otimes I \otimes I \otimes Z  + I \otimes Z \otimes I \otimes I)        \\
        & + 0.045094\ (Y \otimes Y \otimes Y \otimes Y + X \otimes X \otimes Y \otimes Y)         \\
        & + 0.045094\ (Y \otimes Y \otimes X \otimes X + X \otimes X \otimes X \otimes X)         \\
        & + 0.166582\ (X \otimes I \otimes I \otimes Z + I \otimes Z \otimes Z \otimes I)         \\
        & + 0.121488\ (Z \otimes Z \otimes I \otimes I + I \otimes I \otimes Z \otimes Z)
  \end{aligned}
\end{equation}

In order to use the VQE algorithm, we also need to parameterize our quantum circuit and need a suitable initial state.
The VQE algorithm itself has a certain degree of randomness, which does not guarantee that a convergence result can be obtained, and a reasonable selection of the initial state can also accelerate the convergence of the algorithm.
So here, we use UCCSD\cite{uccsd} as an assumption (Ansatz), and the initial state of the circuit is initialized to $\ket{0101}$ using the Hartree-Fock method.

\begin{table}[ht!]
  \vspace*{1em}
  \centering
  \begin{verbatim}
  init_state = HartreeFock(num_spin_orbitals,
    num_particles, converter)
  var_form = UCCSD(
    converter, num_particles,
    num_spin_orbitals, initial_state=init_state)
  \end{verbatim}
  \vspace*{-1.5em}
  \caption{Parameterization of Quantum Circuits}
\end{table}

\begin{figure*}[t]
  \centering
  \includegraphics[scale=0.45]{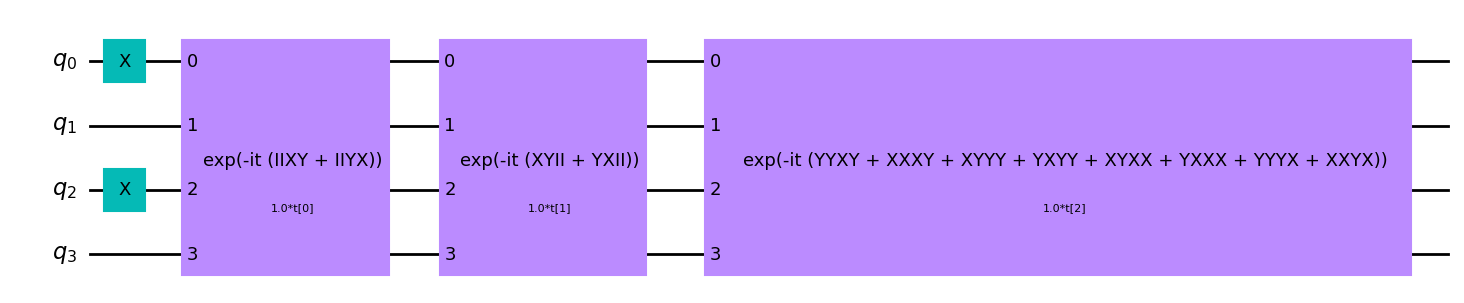}
  \caption{Parameterized Quantum Circuits}
  \label{fig:uucsd}
\end{figure*}

Here, our parameterized quantum circuit can be represented by FIG. \ref{fig:uucsd}.
The three circuit gates connecting four qubits in the figure are circuit gates after packaging.
The mark on the gate indicates which part the gate represents, and $t$ is our parameter.

In this way, the preparatory work is completed, and the next step is to use the VQE algorithm.
Here, we choose to use the IBM Quantum online computing platform provided by IBM Corporation to calculate our circuit.
We can use the \verb|IBMQ| interface to connect to the computing platform.
After selecting the computing device to be used, we can start computing.
Here I choose \textit{ibmq\_jakarta}.

\begin{table}[ht!]
  \vspace*{1em}
  \centering
  \begin{verbatim}
  provider = IBMQ.enable_account(token)
  backend = least_busy(
    provider.backends(
      filters=lambda x:
        x.configuration().n_qubits >= 6
      and not x.configuration().simulator
      and x.status().operational == True))
  print("least busy backend: ", backend)
  \end{verbatim}
  \vspace*{-1.5em}
  \caption{Connect to IBM Quantum and select a free machine}
\end{table}

Similarly, the VQE algorithm has already been implemented in Qiskit, so there is no need to implement it yourself.
At the same time, the VQE algorithm needs to be used in conjunction with classic optimization algorithms, and most of these algorithms are implemented in Qiskit and can be used directly.
Here, for the classical optimization algorithm, considering the situation and accuracy of the quantum computer, the L-BFGS algorithm is chosen to be used.

\begin{table}[ht!]
  \vspace*{1em}
  \centering
  \begin{verbatim}
  optimizer = optimizers.L_BFGS_B(maxiter=20)
  vqe = VQE(var_form, optimizer,
    quantum_instance=backend)
  vqe_calc = vqe.compute_minimum_eigenvalue(
    qubit_op)
  vqe_result =
    problem.interpret(vqe_calc)
      .total_energies[0].real
  \end{verbatim}
  \vspace*{-1.5em}
  \caption{Use the VQE algorithm}
\end{table}

The L-BFGS algorithm, is an optimization algorithm in the family of quasi-Newton methods\cite{quasi}.
This algorithm approximates the Broyden Fletcher Goldfarb Shanno algorithm using limited computer memory.
This algorithm is used to minimize $f(\mathbf{x})$ over an unconstrained vector $\mathbf{x}$, where $f$ is a differentiable function.

As a comparison, we use the result obtained by the built-in ground state energy method in Qiskit as the exact solution, and compare it with the solution obtained by using the VQE algorithm.

\section{Summary}

\begin{figure*}[ht!]
  \centering
  \includegraphics[scale=0.5]{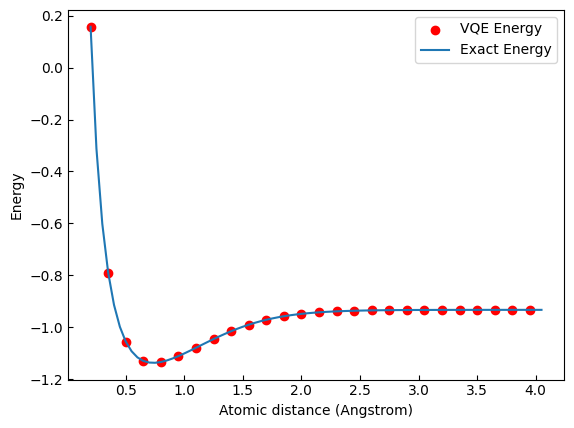}
  \includegraphics[scale=0.5]{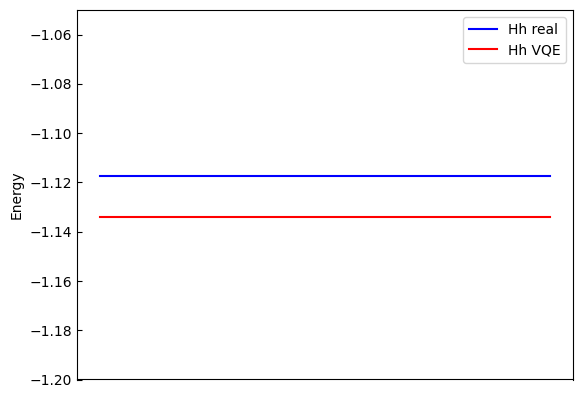}
  \caption{left: Ground state energies of hydrogen molecules at different distances,
    right: Comparison of Actual Energy and Calculated Energy}
  \label{fig:vqeh2}
\end{figure*}

We make the atomic distance start from $0.2$ \r{A}, take $0.05$ \r{A} as the step distance, and continuously calculate the energy ground state of the hydrogen molecular system at different distances, we can get FIG. \ref{fig:vqeh2}.
It can be seen that the results of the VQE algorithm are in good agreement with the exact solution using the numerical method, where the energy units are Hartree.

If compared with the actual data, when the geometric distance is $-0.3625$ and $0.3625$, that is, \verb|dist = 0.725|, the actual data $H_h = -1.117506$, and the data obtained by our VQE algorithm is $ H_h^{\prime} = -1.134167$, the error is about $1.47 \%$.
Although the data difference $0.016661$ is greater than the chemical precision $0.0016$, it is still a relatively accurate solution method.

There may be three sources of error:
\begin{enumerate}
  \item The error caused by assuming that the nucleus is stationary;
  \item There is a certain error in the actual measurement;
  \item The error caused by the randomness of the VQE algorithm itself.
\end{enumerate}

The second error cannot be eliminated, the first error can be reduced using a more accurate model, and the last error may be the most influential.
Especially when running an algorithm on an actual quantum device, due to the limitations of the current quantum device, there is a large background noise, which will affect the coherence of the quantum, thereby affecting the result.

\section*{Acknowledgments}
This work was partially supported by the National Natural Science Foundation of China under Grant Nos. 11875178 and 12005114.
The work of A.W. was supported by the start-up funding from China Three Gorges University.
A.W. is also grateful for the support from the Chutian Scholar Program of Hubei Province.

\bibliography{reference}

\end{document}